\documentclass[prd,aps,showpacs,epsf,floats,onecolumn]{revtex4}%
\usepackage{amssymb}
\usepackage{amsfonts}
\usepackage{amsmath}
\usepackage{graphicx}%
\providecommand{\U}[1]{\protect\rule{.1in}{.1in}}

\begin{document}
\title{\textbf{Maximum Caliber Inference and the Stochastic Ising Model}}
\author{\textbf{Carlo Cafaro}$^{1}$ and \textbf{Sean Alan Ali}$^{2}$}
\affiliation{$^{1}$SUNY Polytechnic Institute, 12203 Albany, New York, USA}
\affiliation{$^{2}$Albany College of Pharmacy and Health Sciences, 12208 Albany, New York, USA}

\begin{abstract}
We investigate the maximum caliber variational principle as an inference
algorithm used to predict dynamical properties of complex nonequilibrium,
stationary, statistical systems in the presence of incomplete information.
Specifically, we maximize the path entropy over discrete time step
trajectories subject to normalization, stationarity, and detailed balance
constraints together with a path-dependent dynamical information constraint
reflecting a given average global behavior of the complex system. A general
expression for the transition probability values associated with the
stationary random Markov processes describing the nonequilibrium stationary
system is computed. By virtue of our analysis, we uncover that a convenient
choice of the dynamical information constraint together with a perturbative
asymptotic\textbf{ }expansion with respect to its corresponding Lagrange
multiplier of the general expression for the transition probability leads to a
formal overlap with the well-known Glauber hyperbolic tangent rule for the
transition probability for the stochastic Ising model in the limit of very
high temperatures of the heat reservoir.

\end{abstract}

\maketitle

\section{Introduction}

A detailed knowledge of microscopic dynamics may be neither necessary nor
sufficient to understand the macroscopic behavior of a complex system. For
example, entropy is a crucially important feature of macrostates that cannot
be determined from microstates. Following the statistical mechanical works of
Gibbs \cite{gibbs74} and inspired by Shannon's advances in information theory
\cite{shannon48}, Jaynes is the undisputed pioneer in the use of entropy for
optimal information processing leading to reliable macroscopic predictions in
the presence of incomplete information \cite{jaynes57A, jaynes57B, jaynes80,
jaynes85, jaynes86}. We point out that information processing is considered
optimal when it takes into account all available knowledge of the microphysics
as well as all the available macroscopic data. No arbitrary assumptions beyond
that are introduced.

\subsection{Jaynes's MaxEnt}

Jaynes's MaxEnt is an inductive method of reasoning for making predictions
about \emph{equilibrium properties} of macroscopic phenomena in the presence
of incomplete information \cite{jaynes57A}. The only type of initial
information allowed is values of quantities which are observed to be constant
in time. In synthesis, MaxEnt is a \emph{state-based} variational method of
information-theoretic nature aiming at inferring macroscopic (conserved)
properties of complex systems at equilibrium in the presence of limited
information about the microscopic nature of the phenomena being investigated.
Macroscopic predictions generated by the MaxEnt inference algorithm are
related to the experimental behavior of actual physical systems only when, and
to the extent that, they lead to sharply peaked probability distributions.
Furthermore, if it occurs that there is experimental evidence that a given
MaxEnt prediction is incorrect, one should reasonably conclude that the
enumeration of the possible microstates suggested by our knowledge of the laws
of physics was not correctly given. The MaxEnt inference algorithm should be
applied again by modifying the set of input information constraints
corresponding to the newly updated enumeration of the microstates of the
system. Indeed, failure of the MaxEnt algorithm could be regarded as more
valuable than its success since this can lead to fundamental advances in
science \cite{jaynes86}. For instance, the failures of classical statistical
mechanics that are\textbf{ }rectified by quantum theory constitute valid
examples of such a possibility. Despite a few weak technical aspects of
Jaynes's approach \cite{rondoni14}, it is unjustified to state that physics
becomes irrelevant in the MaxEnt formalism. As pointed out earlier, failures
of the MaxEnt inference algorithm are ascribed to its physical rather than
statistical aspects. The MaxEnt inference is merely an algorithm, a messenger
\cite{dewar09}. The MaxEnt inference algorithm ensures the objectivity of its
predictions but does not claim deductive certainty for them. Ultimately,
conclusive successes and failures belong to physics. Having said that, we also
have to point out that the application of MaxEnt is not limited to statistical
mechanics. Practical applications of the MaxEnt algorithm include, but are not
limited to, image reconstruction in radio and x-ray astronomy \cite{gull78,
skilling84}, image reconstruction in medical tomography \cite{minerbo79},
x-ray crystallography \cite{collins82}, molecular biology \cite{bryan83},
nuclear magnetic resonance spectroscopy \cite{sibisi84}, and the collective
dynamics of a population of neurons \cite{mora15}. From a more theoretical
perspective, several applications of the MaxEnt algorithm appear in the
characterization of complexity of natural phenomena \cite{carlopd07,
carlochaos,carlopd11,henry16}, energy levels statistics \cite{carlomplb08,
carlopa08}, quantum entanglement \cite{carlopla, carlopa}, and ferromagnetic
materials \cite{carlo16}.

\subsection{Jaynes's MaxCal}

The extension of Jaynes's MaxEnt principle to nonequilibrium statistical
mechanical phenomena is known as Jaynes's MaxCal principle \cite{jaynes80,
jaynes85}. It constitutes an inductive method of reasoning for making
predictions of \emph{nonequilibrium properties} of macroscopic phenomena in
the presence of incomplete information. The type of initial information
allowed is extended to values of quantities which are observed to be
nonconstant in time. Unlike MaxEnt, MaxCal is a \emph{path-based} variational
method of information-theoretic nature aiming at inferring macroscopic
(dynamical) properties of complex systems out of equilibrium in the presence
of partial knowledge about the microscopic nature of the phenomena of
interest. From an applied perspective, it is found that MaxCal is especially
suitable for solving problems involving small systems that are out of
equilibrium where fluctuations become important. Systems of this type appear
naturally in biology and nanotechnology \cite{ghosh06, seita07, wu09}.
Specifically, in Refs. \cite{ghosh06, seita07} MaxCal was applied to derive
several flux distributions in diffusive systems: Fick's law of particle
transport, Fourier's law of heat transport, and Newton's viscosity law of
momentum transport. In Ref. \cite{wu09}, instead, MaxCal was experimentally
tested in the study of a single colloidal particle transitioning between two
energy wells. In recent years, MaxCal has also found application to the
dynamics of collectively moving animal groups \cite{cavagna14}. Recent years
have also witnessed a significant number of more theoretically oriented
applications of the MaxCal inference algorithm \cite{dewar09, jtd94, dewar03,
gonzalez13, davis15}. In particular, MaxCal has been employed to derive master
equations \cite{stock08, lee12} and Markov processes \cite{ge12}, it has been
challenged to make inferences in the realm of non-Markovian dynamics
\cite{otten10}, it has been used to recover classical Newtonian equations of
motion \cite{gonzalez13} and the Fokker-Planck equation \cite{davis15} and,
finally, it has been proposed as a general variational principle for
nonequilibrium statistical mechanics \cite{hazoglou15}. For an extensive
review on the MaxCal formalism, we refer to \cite{presse13}. We emphasize that
when taking into consideration nonequilibrium scenarios, Jaynes was concerned
with continuous paths satisfying deterministic Hamiltonian equations of motion
\cite{jaynes80, jaynes85} while Filyukov and Karpov focused their attention on
systems having discrete dynamical states where trajectories are composed of
discrete time steps \cite{karpov1, karpov2, karpov3}. Specifically, they
assumed that the evolution of the system may be specified by a Markov chain
with discrete times and a finite number of states. In particular, they noticed
that in addition to the state probabilities, path probabilities were required
to study nonequilibrium processes since it was important to know the rates of
transition from one state to another \cite{karpov1, karpov2}. This discrete
line of investigation is generating very interesting findings. In Ref.
\cite{dixit14}, the MaxCal inference procedure was used to infer the
transition probabilities in a stationary Markov process given the knowledge of
both the stationary-state populations and an average global dynamical
quantity. Specifically, Dixit and Dill showed that their work yields the
correct values of dynamical quantities (normalized occupancy autocorrelation)
in an example of molecular dynamics simulations of a water solvation shell
around a single water molecule (the average fluctuation of the number of
molecules in the hydration shell was assumed known). In Ref. \cite{dixit15A},
the MaxCal inference procedure was used to infer both the stationary-state
distributions and the transition probabilities in a stationary Markov process
in the presence of both state- and path-dependent information constraints.
Specifically, Dixit considered a particle diffusing on a two-dimensional
square lattice where the external potential field was used as the average
global dynamical quantity in the path entropy maximization procedure. In
\cite{dixit15B}, MaxCal was applied to network science in order to infer
transition rates between nodes of a network in the presence of partial
knowledge in the form of both state- and path-dependent information. In
particular, when the average global dynamical quantity known was represented
by the mean jump rate, it was uncovered that the transition rates $\omega
_{ij}$ exhibit a square-root dependence on the stationary state populations at
nodes $i$ and $j$.

In principle, the Liouville equation is the conventional starting point for
the description of time-dependent statistical systems \cite{huangbook}.
However, in most spin systems, the lack of complete knowledge leads to the
absence of an explicit form of the interaction Hamiltonian between the spins
and the heat reservoir \cite{huang73}. This fact remains true even in the case
of the linear Ising chain, the simplest Ising model \cite{ising}. For this
reason, the Liouville equation is not useful whereas the method of the master
equation allows one to overcome this technical difficulty. Indeed, within this
latter method, it is only necessary to know the Hamiltonian of the spin system
while the interaction that causes the time transition is assumed to be
stochastic. This line of reasoning leads to the so-called Glauber stochastic
Ising model \cite{glauber63}. As pointed out by Glauber himself in Ref.
\cite{glauber63}, the functional form of the transition probabilities proposed
by him was motivated by simplicity requirements rather than generality conditions.

In this article, motivated by the considerations presented in Ref.
\cite{glauber63} and especially inspired by the works presented in Ref.
\cite{dixit14, dixit15A, dixit15B}, we maximize the path entropy over discrete
time step trajectories subject to normalization, stationarity, and detailed
balance constraints together with a path-dependent dynamical information
constraint reflecting a given average global behavior of the complex spin
system being considered. We compute a general expression for the transition
probability values associated with the stationary random Markov processes
describing the nonequilibrium stationary system. Finally, we uncover that a
convenient choice of the dynamical information constraint together with a
perturbative expansion with respect to the corresponding Lagrange multiplier
of the general expression for the transition probability leads to a formal
overlap with the well-known Glauber hyperbolic tangent rule for the transition
probability for the stochastic Ising model \cite{penrose91} in the limit of
very high temperatures of the heat reservoir.

The layout of this article is as follows. In Sec. II, we briefly describe the
stochastic Ising model as proposed by Glauber. In particular, we recast
Glauber's hyperbolic tangent rule in a form more convenient for our analysis.
In Sec. III, we present a synthetic picture of the philosophy underlying the
MaxCal formalism. We focus our attention on the statistical inference of
nonequilibrium properties of complex systems characterized by discrete
dynamical states where paths are composed of discrete time steps. Section IV
is divided into two parts. In the first part, we show a detailed computation
of the path entropy maximization procedure in the presence of various
information constraints, including a path ensemble average of the product of
two neighboring spin values. In the second part, we formally compare the
outcome of our computation with Glauber's transition probability expression in
two limiting cases. Finally, our conclusions are reported in Sec. V.

\section{The Stochastic Ising Model}

As pointed out in the introduction, the Liouville equation would be the
standard starting point for the description of time-dependent statistical
systems. However, in most cases, an explicit form of the interaction
Hamiltonian between the spins and the heat reservoir is not fully known. For
this reason, the Liouville equation is not useful in such a case and the
method of the master equation allows one to overcome this technical
difficulty. Indeed, within this method, it is only necessary to know the
Hamiltonian of the spin system while the interaction that causes the time
transition is assumed to be stochastic. The master equation is given by
\cite{binder05},%
\begin{equation}
\frac{\partial p_{i}\left(  t\right)  }{\partial t}=-%
{\displaystyle\sum\limits_{i\neq j}}
\left[  p_{i}\left(  t\right)  \omega_{i\rightarrow j}-p_{j}\left(  t\right)
\omega_{j\rightarrow i}\right]  \text{,}%
\end{equation}
where $p_{i}\left(  t\right)  $ is the probability of the system being in the
state $i$ at time $t$, while $\omega_{i\rightarrow j}$ is the transition rate
for $i\rightarrow j$. At equilibrium,%
\begin{equation}
\frac{\partial p_{i}\left(  t\right)  }{\partial t}=0\text{,}%
\end{equation}
that is,%
\begin{equation}
\frac{p_{i}\left(  t\right)  }{p_{j}\left(  t\right)  }=\frac{\omega
_{j\rightarrow i}}{\omega_{i\rightarrow j}}\text{.} \label{detailed}%
\end{equation}
Equation (\ref{detailed}) is known as the detailed balance condition. Observe
that, in explicit analogy with equilibrium statistical mechanics, the
probability of the $i$th state occurring in a classical system is given by,%
\begin{equation}
p_{i}\left(  t\right)  =\frac{1}{\mathcal{Z}}e^{-\frac{E_{i}}{k_{\text{B}}T}%
}\text{,} \label{salva}%
\end{equation}
where $\mathcal{Z}$ is the partition function and $E_{i}$ is the energy of the
spin in the $i$th state. Usually, this probability is only approximately known
due to the denominator \cite{binder05}.

In the absence of any external magnetic field and in the working hypothesis
that each spin $\sigma_{i}$ is coupled through the transition probabilities
$\omega_{i}\left(  \sigma_{i}\rightarrow-\sigma_{i}\right)  $ to only its
nearest-neighbor spins $\sigma_{i-1}$ and $\sigma_{i+1}$, and motivated by
both simplicity and symmetry arguments, Glauber considered a stochastic Ising
model described by a master equation in which the transition probabilities
were given by \cite{glauber63},%
\begin{equation}
\omega_{i}\left(  \sigma_{i}\rightarrow-\sigma_{i}\right)  =\frac{\Gamma}%
{2}\left\{  1-\frac{1}{2}\gamma\sigma_{i}\left(  \sigma_{i-1}+\sigma
_{i+1}\right)  \right\}  \text{.} \label{uno}%
\end{equation}
In Eq. (\ref{uno}), $\Gamma$ is a parameter that characterizes the time scale
on which all transitions take place, $\Gamma/2$ denotes the rate per unit time
at which the particle makes transitions from either state to the opposite, and
$\gamma$ is a parameter that describes the tendency of spins toward alignment.
The explicit expression for $\gamma$ in Eq. (\ref{uno}) can be obtained by
imposing the detailed balancing condition at equilibrium at temperature $T$
for the Ising model,%
\begin{equation}
\frac{p_{i}\left(  -\sigma_{i}\right)  }{p_{i}\left(  \sigma_{i}\right)
}=\frac{\omega_{i}\left(  \sigma_{i}\rightarrow-\sigma_{i}\right)  }%
{\omega_{i}\left(  -\sigma_{i}\rightarrow\sigma_{i}\right)  }\text{.}
\label{due}%
\end{equation}
The quantity $p_{i}\left(  \sigma_{i}\right)  $ in Eq. (\ref{due}) denotes the
probability that the $i$th spin will assume the value $\sigma_{i}$ and is
proportional to the Maxwell-Boltzmann factor,%
\begin{equation}
p_{i}\left(  \sigma_{i}\right)  \propto\exp\left[  -\frac{J}{k_{\text{B}}%
T}\sigma_{i}\left(  \sigma_{i-1}+\sigma_{i+1}\right)  \right]  \text{,}
\label{tre}%
\end{equation}
where $J$ is the exchange coupling constant and $k_{\text{B}}$ is the
Boltzmann constant. Substituting Eqs. (\ref{uno}) and (\ref{tre}) into Eq.
(\ref{due}), we obtain%
\begin{equation}
\gamma=\gamma\left(  k_{\text{B}}\text{, }J\text{, }T\right)  =\tanh\left(
\frac{2J}{k_{\text{B}}T}\right)  \text{.} \label{quattro}%
\end{equation}
Substituting Eq. (\ref{quattro}) into Eq. (\ref{uno}) and using the fact that
$\sigma_{i}=\pm1$ together with the point symmetry of the hyperbolic tangent
function, we find%
\begin{equation}
\omega_{i}\left(  \sigma_{i}\rightarrow-\sigma_{i}\right)  =\frac{\Gamma}%
{2}\left[  1-\sigma_{i}\tanh\left(  \beta h_{i}\right)  \right]  \text{,}
\label{gg}%
\end{equation}
where $\beta\overset{\text{def}}{=}\frac{1}{k_{\text{B}}T}$ and $h_{i}$
denotes a local magnetic field defined as,%
\begin{equation}
h_{i}\overset{\text{def}}{=}J\left(  \sigma_{i-1}+\sigma_{i+1}\right)
\text{.} \label{hi}%
\end{equation}
Equation (\ref{gg}) is the so-called Glauber hyperbolic tangent rule. For
future use, we recast this equation in a more convenient form. Using again the
fact that $\sigma_{i}=\pm1$ and exploiting the point symmetry of the
hyperbolic tangent function, it follows that%
\begin{equation}
\sigma_{i}\tanh\left(  \beta h_{i}\right)  =\tanh\left(  \beta\sigma_{i}%
h_{i}\right)  \text{.} \label{g5}%
\end{equation}
Observe that the energy difference $\Delta E$ between a proposed new
microstate $j$ and an old microstate $i$ is given by \cite{janke12},%
\begin{equation}
\Delta E\overset{\text{def}}{=}E_{j}-E_{i}=2\sigma_{i}h_{i}\text{.} \label{g6}%
\end{equation}
Finally, inserting Eqs. (\ref{g5}) and (\ref{g6}) into Eq. (\ref{gg}), after
some algebra we obtain%
\begin{equation}
\omega_{ij}=\frac{\Gamma}{2}\left[  1-\tanh\left(  \frac{\beta\Delta E}%
{2}\right)  \right]  \text{,} \label{g7}%
\end{equation}
that is,%
\begin{equation}
\left[  \omega_{ij}\right]  _{\text{Glauber}}=\Gamma\frac{e^{-\frac
{\beta\Delta E}{2}}}{e^{\frac{\beta\Delta E}{2}}+e^{-\frac{\beta\Delta E}{2}}%
}\text{.} \label{omega}%
\end{equation}
For the sake of clarity, we remark that Glauber pointed out in Ref.
\cite{glauber63} that in the working hypothesis of nearest-neighbor coupling
among spins, the functional form of the transition probability $\omega_{ij}$
that leads to the same equilibrium state as the Ising model is not unique. The
specific functional form of $\omega_{ij}$ in Eq. (\ref{uno}) as employed by
Glauber in the master equation formalism was selected for simplifying the
equations to handle (for instance, equations describing the spin expectation
values) rather than satisfying a fundamental physical requirement. More
specifically, the manner in which $\omega_{ij}$ depends on neighboring spin
values chosen by Glauber was dictated by the necessity of describing a
tendency for each spin to align itself parallel to its nearest neighbors. In
summary, the assumption that $\omega_{ij}$ depends symmetrically on the two
neighboring spins $\sigma_{i-1}$ and $\sigma_{i+1}$ as well as $\sigma_{i}$ is
a clever \emph{ad hoc} assumption adopted by Glauber.

The expression for the transition probability $\omega_{ij}$ in Eq.
(\ref{omega}) will be useful in the remainder of the article.

\section{The MaxCal Formalism}

In the MaxCal formalism \cite{jaynes80, jaynes85, karpov1, karpov2}, the path
entropy to be maximized can be defined as \cite{karpov1, karpov2},%
\begin{equation}
H\left(  \left\{  p\left(  C\right)  \right\}  \right)  \overset{\text{def}%
}{=}-\sum_{\left\{  C\right\}  }p\left(  C\right)  \ln\left[  p\left(
C\right)  \right]  \text{,} \label{PE}%
\end{equation}
where $p\left(  C\right)  $ denotes the probability that the dynamical process
follows the path $C$. As mentioned in the introduction, we recall that Jaynes
was concerned with continuous paths $C$ satisfying deterministic Hamiltonian
equations of motion \cite{jaynes80, jaynes85}. However, for systems having
continuous dynamical states, it is computationally hard to compute a path
ensemble using the microscopic dynamics. Rather than considering continuous
paths, Filyukov and Karpov focused their attention on systems having discrete
dynamical states where trajectories are composed of discrete time steps
\cite{karpov1, karpov2}. Specifically, they assumed that the evolution of the
system may be specified by a Markov chain with discrete times and a finite
number of states. The trajectory $C_{T}$ of a Markov chain of length $T$ is
described by the sequence of states,%
\begin{equation}
C_{T}=i_{0}i_{1}\cdot\cdot\cdot i_{T-1}i_{T}\text{.}%
\end{equation}
Assuming a stationary first-order Markov process \cite{kampen81, cover}, the
probability $p\left(  C_{T}\right)  $ of the trajectory is given by,%
\begin{equation}
p\left(  C_{T}\right)  =p_{i_{0}}\omega_{i_{0}i_{1}}\cdot\cdot\cdot
\omega_{i_{T-1}i_{T}}\text{,}%
\end{equation}
where $\omega_{ij}=\omega_{ij}\left(  \tau\right)  $ denotes the conditional
probability of a transition on the time interval $\tau$ from the state $i$ to
the state $j$ while $p_{i_{0}}$ is the single state $i_{0}$ occupation
probability. Note that the conditional transition probabilities $\omega_{ij}$
and the stationary single state probabilities $p_{i}$ satisfy the
normalization constraints,%
\begin{equation}
\sum_{j}\omega_{ij}=1\text{, and }\sum_{i}p_{i}=1\text{,}%
\end{equation}
respectively. Furthermore, the stationarity of the chain is encoded in the
following constraint,%
\begin{equation}%
{\displaystyle\sum\limits_{i}}
p_{i}\omega_{ij}=p_{j}\text{.}%
\end{equation}
It can be shown that for a sufficiently long ergodic chain, that is to say a
chain for which $T$ approaches infinity and $\omega_{ij}>0$ (all states are
connected by a nonzero transition probability), the path entropy in Eq.
(\ref{PE}) can be approximated by (for further details, see Refs.
\cite{karpov1, karpov2, presse13})%
\begin{equation}
H\left(  \left\{  p\left(  C\right)  \right\}  \right)  \approx H\left(
T\right)  \overset{\text{def}}{=}TH\left(  1\right)  \text{,}%
\end{equation}
where $H\left(  1\right)  $ denotes the path entropy per step defined as,%
\begin{equation}
H\left(  1\right)  \overset{\text{def}}{=}-\sum_{i,j}p_{i}\omega_{ij}%
\ln\left[  \omega_{ij}\right]  \text{.} \label{h1}%
\end{equation}
Observe that the path entropy per step reduces to the ordinary entropy of
equilibrium statistical mechanics when $\omega_{ij}=p_{j}$, that is to say
when an instant equilibration condition is achieved.

The path entropy in Eq. (\ref{PE}) is also known as the \emph{caliber}, a
cross sectional area of a tube that, in part, quantifies the flow in a dynamic
process. By maximizing the caliber subject to all available constraints
(normalization conditions, dynamical averages, etc.), MaxCal yields the least
biased probability $p\left(  C\right)  $ for the set of microscopic
trajectories $\left\{  C\right\}  $ consistent with the observed information
constraints. Specifically, given the knowledge of all possible microscopic
trajectories explored in a specific interval of time by a system, MaxCal seeks
to construct a weighted ensemble of microscopic trajectories consistent with
the constrained averages obtained by measuring a small (much smaller than the
number of known microscopic trajectories) number of dynamical quantities (for
instance, average microscopic fluxes). In turn, this weighted ensemble of
microscopic trajectories determines the time evolution of all time-dependent
observables of the system. In summary, in analogy to MaxEnt, macroscopic
quantities are computed in terms of derivatives of a dynamical partition
function defining the normalization factor of the least biased probability
$p\left(  C\right)  $.

\section{Path Entropy Maximization and Neighboring Spin Values}

In what follows, we assume that the path entropy per unit time to be maximized
is given by the caliber in Eq. (\ref{h1}),%
\begin{equation}
\mathcal{C}\overset{\text{def}}{=}-\sum_{i,j}p_{i}\omega_{ij}\ln\left[
\omega_{ij}\right]  \text{.} \label{caliber}%
\end{equation}

\subsection{The Explicit Computation}

The first information constraint that we impose is the transition probability
normalization constraint,%
\begin{equation}
\sum_{j}\omega_{ij}=1\text{, }\forall i
\end{equation}
that is,%
\begin{equation}
\sum_{j}p_{i}\omega_{ij}=p_{i}\text{, }\forall i\text{.} \label{c1}%
\end{equation}
This constraint describes the fact that from the state $i$ at time $t$, the
system has to transition to some state $j$ at time $t+\delta t$. The second
information constraint that we consider is the stationary state probability
normalization constraint given by,%
\begin{equation}
\sum_{i,j}p_{i}\omega_{ij}=1 \label{c2}%
\end{equation}
that is,%
\begin{equation}
\sum_{j}p_{j}=1\text{.}%
\end{equation}
The third constraint is the stationarity constraint described in terms of the
following constraining relation,%
\begin{equation}
\sum_{i}p_{i}\omega_{ij}=p_{j}\text{, }\forall j\text{.} \label{c3}%
\end{equation}
The constraint in Eq. (\ref{c3}) describes the fact that a system in state $j$
at time $t+\delta t$ comes from one of the states $i$ at time $t$. Before
introducing the fourth constraint, we observe that within the MaxCal formalism
a global constraint\textbf{ }$\left\langle \sigma\left(  t\right)
\right\rangle $\textbf{ }is defined in terms of a path ensemble average of a
dynamical quantity\textbf{ }$\sigma\left(  t\right)  $\textbf{ }that depends
on both the initial and final states $i$\textbf{ }and\textbf{ }$j$\textbf{,
}respectively. Specifically \cite{karpov2, dixit14}\textbf{,}%
\begin{equation}
\left\langle \sigma\left(  t\right)  \right\rangle \overset{\text{def}}{=}%
\sum_{\left\{  C\right\}  }p\left(  C\right)  \left\langle \sigma\right\rangle
_{C}=%
{\displaystyle\sum\limits_{i,j}}
p_{i}\omega_{ij}\sigma_{ij}\text{,}%
\end{equation}
where\textbf{ }$\left\langle \sigma\right\rangle _{C}$\textbf{ }denotes the
average of\textbf{ }$\sigma$\textbf{ }over a steady state path\textbf{
}$C\overset{\text{def}}{=}\cdot\cdot\cdot a\rightarrow b\rightarrow
c\rightarrow d\cdot\cdot\cdot$\textbf{ }of length\textbf{ }$T$\textbf{,}%
\begin{equation}
\left\langle \sigma\right\rangle _{C}\overset{\text{def}}{=}\frac{1}{T}\left(
\cdot\cdot\cdot\sigma_{ab}+\sigma_{bc}+\sigma_{cd}+\cdot\cdot\cdot\right)
\text{.}%
\end{equation}
We assume that the fourth constraint is described in terms of a path ensemble
average of a dynamical quantity\textbf{ }$\sigma=\sigma\left(  t\right)
$\textbf{ }specified by the product of two spin values\textbf{ }$\sigma
_{i}\left(  t\right)  $\textbf{ }and\textbf{ }$\sigma_{j}\left(  t\right)  $.
This constraint is given by,%
\begin{equation}
\left\langle \sigma\left(  t\right)  \right\rangle =%
{\displaystyle\sum\limits_{i,j}}
p_{i}\omega_{ij}\sigma_{ij}\text{,} \label{c4}%
\end{equation}
where we can generally define\textbf{ }$\sigma_{ij}\left(  t\right)
\overset{\text{def}}{=}\sigma_{i}\left(  t\right)  \sigma_{j}\left(  t\right)
$\textbf{. }Since each spin\textbf{ }$\sigma_{i}$\textbf{ }is coupled to only
its nearest-neighbor spins\textbf{ }$\sigma_{i-1}$\textbf{ }and\textbf{
}$\sigma_{i+1}$\textbf{ }in Glauber's stochastic Ising model, we make the
following clarifying remark. For nearest-neighbor pairs of spins\textbf{
}$\sigma_{i}$ and $\sigma_{j}$\textbf{ }with\textbf{ }$j\in\left\{
i\pm1\right\}  $\textbf{ }we consider\textbf{ }$\sigma_{ij}=\sigma_{i}%
\sigma_{i+1}$\textbf{, }$\sigma_{ji}=\sigma_{i-1}\sigma_{i}$\textbf{
}and\textbf{, }$\sigma_{ij}+\sigma_{ji}=\sigma_{i}\left(  \sigma_{i-1}%
+\sigma_{i+1}\right)  \propto\Delta E$\textbf{, }where\textbf{ }$\Delta E$ is
the energy\textbf{ }difference between an old and a new microstate of the spin
system as presented in Eq. (\ref{g6}). Finally, the fifth condition we impose
is the detailed balance constraint given by,%
\begin{equation}
p_{i}\omega_{ij}=p_{j}\omega_{ji}\text{.} \label{c5}%
\end{equation}
Given the path entropy per unit time in Eq. (\ref{caliber}) and the
information constraints in Eqs. (\ref{c1}), (\ref{c2}), (\ref{c3}),
(\ref{c4}), and (\ref{c5}), the caliber to be maximized becomes,%
\begin{align}
\mathcal{C}=-  &  \sum_{i,j}p_{i}\omega_{ij}\ln\left[  \omega_{ij}\right]
+\sum_{i}\alpha_{i}\left(  \sum_{j}p_{i}\omega_{ij}-p_{i}\right)
+\beta\left(  \sum_{i,j}p_{i}\omega_{ij}-1\right)  +\nonumber\\
& \nonumber\\
&  +\sum_{j}\gamma_{j}\left(  \sum_{i}p_{i}\omega_{ij}-p_{j}\right)
+\delta\left(
{\displaystyle\sum\limits_{i,j}}
p_{i}\omega_{ij}\sigma_{ij}-\left\langle \sigma\right\rangle \right)  +%
{\displaystyle\sum\limits_{i,j}}
\xi_{ij}\left(  p_{i}\omega_{ij}-p_{j}\omega_{ji}\right)  \text{,}%
\end{align}
where $\alpha_{i}$, $\beta$, $\gamma_{j}$, $\delta$, and $\xi_{ij}$ are
Lagrange multipliers. Note that the variation of the caliber $\mathcal{C}$
with respect to the (unknown) stationary state probability $p_{i}$ and the
transition probability $\omega_{ij}$ is given by $\delta\mathcal{C}$,%
\begin{equation}
\delta\mathcal{C}=\frac{\delta\mathcal{C}}{\delta\omega_{ij}}\delta\omega
_{ij}+\frac{\delta\mathcal{C}}{\delta p_{i}}\delta p_{i}\text{.}
\label{deltac}%
\end{equation}
For the sake of completeness, we emphasize at this juncture that there are
scenarios in which variations of functionals appear with respect to the
Lagrange multipliers. For instance, in the presence of space-time dependent
information constraints in transport theory, the covariance functions
$\mathcal{K}_{ij}$ are expressed in terms of the second functional derivative
of $\ln\left(  \mathcal{Z}\right)  $ as follows \cite{jaynes85},%
\begin{equation}
\mathcal{K}_{ij}\left(  x,t;x^{\prime},t^{\prime}\right)  \overset{\text{def}%
}{=}\frac{\delta^{2}\left[  \ln\left(  \mathcal{Z}\right)  \right]  }%
{\delta\lambda_{i}\left(  x,t\right)  \delta\lambda_{j}\left(  x^{\prime
},t^{\prime}\right)  }\text{,}%
\end{equation}
where $\mathcal{Z}=\mathcal{Z}\left(  \left\{  \lambda_{i}\left(  x,t\right)
\right\}  \right)  $ denotes the partition functional while $\lambda_{i}$ are
the Lagrange multipliers with $1\leq i,j\leq m$ and $m$ denoting the
cardinality of the information constraints being chosen. Within the MaxCal
formalism, Lagrange multipliers are generated by the first functional
derivatives of the caliber while higher derivatives of the partition function
lead to higher moments of the observables. The stationarity of $\delta
\mathcal{C}$ in Eq. (\ref{deltac}) requires that both $\frac{\delta
\mathcal{C}}{\delta\omega_{ij}}$ and $\frac{\delta\mathcal{C}}{\delta p_{i}}$
must simultaneously vanish. Let us observe that, after some algebra,
$\frac{\delta\mathcal{C}}{\delta\omega_{ij}}\delta\omega_{ij}$ is given by%
\begin{equation}
\frac{\delta\mathcal{C}}{\delta\omega_{ij}}\delta\omega_{ij}=-\left\{
\sum_{i,j}\left[  p_{i}\ln\omega_{ij}+p_{i}-\alpha_{i}p_{i}-\beta p_{i}%
-\gamma_{j}p_{i}-\delta p_{i}\sigma_{ij}-p_{i}\xi_{ij}+p_{i}\xi_{ji}\right]
\delta\omega_{ij}\right\}  \text{,}%
\end{equation}
that is,
\begin{equation}
\omega_{ij}=e^{\alpha_{i}+\beta+\gamma_{j}+\delta\sigma_{ij}+\left(  \xi
_{ij}-\xi_{ji}\right)  -1}\text{.} \label{e1}%
\end{equation}
Furthermore, let us notice that $\frac{\delta\mathcal{C}}{\delta p_{i}}\delta
p_{i}$ becomes, after some algebra%
\begin{equation}
\frac{\delta\mathcal{C}}{\delta p_{i}}\delta p_{i}=\left\{
\begin{array}
[c]{c}%
-\sum_{j}\omega_{ij}\ln\omega_{ij}+\alpha_{i}\sum_{j}\omega_{ij}-\alpha
_{i}+\beta\sum_{j}\omega_{ij}+\\
\\
+\sum_{j}\gamma_{j}\omega_{ij}-\gamma_{j}+\delta\sum_{j}\omega_{ij}\sigma
_{ij}+\sum_{j}\omega_{ij}\left(  \xi_{ij}-\xi_{ji}\right)
\end{array}
\right\}  \delta p_{i}\text{,}%
\end{equation}
that is,%
\begin{align}
\sum_{j}\omega_{ij}\ln\omega_{ij}  &  =+\alpha_{i}\sum_{j}\omega_{ij}%
-\alpha_{i}+\beta\sum_{j}\omega_{ij}+\sum_{j}\gamma_{j}\omega_{ij}-\gamma
_{j}+\nonumber\\
& \nonumber\\
&  +\delta\sum_{j}\omega_{ij}\sigma_{ij}+\sum_{j}\omega_{ij}\left(  \xi
_{ij}-\xi_{ji}\right)  \text{.} \label{e2}%
\end{align}
Substituting Eq. (\ref{e1}) into Eq. (\ref{e2}), after some manipulations, we
obtain the following relation between the Lagrange multipliers $\alpha_{i}$
and $\gamma_{i}$,
\begin{equation}
\alpha_{i}+\gamma_{i}=1\text{, }\forall i\text{.} \label{e3}%
\end{equation}
For the sake of notational simplicity, let us relabel the Lagrange multipliers
as follows,%
\begin{equation}
A_{i}\overset{\text{def}}{=}e^{-\alpha_{i}}\text{, }eA_{j}\overset{\text{def}%
}{=}e^{\gamma_{j}}\text{, }B\overset{\text{def}}{=}e^{-\beta}\text{, and
}E_{ij}\overset{\text{def}}{=}e^{\xi_{ij}-\xi_{ji}}\text{,}%
\end{equation}
where $e$ denotes the Neper constant. Then, using Eqs. (\ref{e1}) and
(\ref{e3}), the transition probability $\omega_{ij}$ becomes%
\begin{equation}
\omega_{ij}=\frac{1}{B}\frac{A_{j}}{A_{i}}e^{\delta\sigma_{ij}}E_{ij}\text{.}
\label{e4}%
\end{equation}
The quantity $E_{ij}$ can be obtained by imposing the detailed balance
constraint in Eq. (\ref{c5}). After some algebra, it is found that,%
\begin{equation}
\frac{\omega_{ij}}{\omega_{ji}}=\frac{p_{j}}{p_{i}}=\left(  \frac{A_{j}}%
{A_{i}}\right)  ^{2}e^{\delta\left(  \sigma_{ij}-\sigma_{ji}\right)  }%
E_{ij}^{2}\text{,}%
\end{equation}
that is,%
\begin{equation}
E_{ij}=\sqrt{\frac{p_{j}}{p_{i}}}\frac{A_{i}}{A_{j}}e^{-\frac{\delta}%
{2}\left(  \sigma_{ij}-\sigma_{ji}\right)  }\text{.} \label{e51}%
\end{equation}
Finally, combining Eqs. (\ref{e4}) and (\ref{e51}), we obtain%
\begin{equation}
\left[  \omega_{ij}\right]  _{\text{Max-Cal}}=\mathcal{N}\sqrt{\frac{p_{j}%
}{p_{i}}}e^{-\frac{\gamma}{2}\left(  \sigma_{ij}+\sigma_{ji}\right)  }\text{,}%
\end{equation}
where $\mathcal{N}$ and $\gamma$ are defined as,%
\begin{equation}
\mathcal{N}\overset{\text{def}}{=}\frac{1}{B}\text{, and }\gamma
\overset{\text{def}}{=}-\delta\text{,}%
\end{equation}
respectively. The Lagrange multiplier $B=$ $\mathcal{N}^{-1}$ can be obtained
by imposing the normalization condition,%
\begin{equation}%
{\displaystyle\sum\limits_{j}}
\omega_{ij}=1\text{,}%
\end{equation}
that is,%
\begin{equation}%
{\displaystyle\sum\limits_{j}}
W_{ij}\phi_{j}=B\phi_{i}\text{,}%
\end{equation}
with $\phi_{i}\overset{\text{def}}{=}\sqrt{p_{i}}$ and $W_{ij}\overset
{\text{def}}{=}e^{-\frac{\gamma}{2}\left(  \sigma_{ij}+\sigma_{ji}\right)  }$.
Furthermore, the Lagrange multiplier $\delta$ can be obtained by imposing the
path ensemble average of the dynamical variable $\sigma\left(  t\right)  $ in
Eq. (\ref{c4}). After some algebraic manipulations, we finally determine%
\begin{equation}
\left[  \omega_{ij}\right]  _{\text{Max-Cal}}=\mathcal{N}\sqrt{\frac{p_{j}%
}{p_{i}}}\cosh\left[  \frac{\gamma}{2}\left(  \sigma_{ij}+\sigma_{ji}\right)
\right]  \left\{  1-\tanh\left[  \frac{\gamma}{2}\left(  \sigma_{ij}%
+\sigma_{ji}\right)  \right]  \right\}  \text{,} \label{fifi}%
\end{equation}
where $\sigma_{ij}\left(  t\right)  \overset{\text{def}}{=}\sigma_{i}\left(
t\right)  \sigma_{j}\left(  t\right)  $. The functional form of the transition
probabilities in Eq. (\ref{fifi}) is the one inferred by the MaxCal inference
algorithm given the information constraints in Eqs. (\ref{c1}), (\ref{c2}),
(\ref{c3}), (\ref{c4}), and (\ref{c5}).

\subsection{Statistical mechanical remarks}

In what follows, several statistical mechanical remarks are presented. First,
within statistical mechanics, the common wisdom is that high temperatures lead
to decay of correlations \cite{gross79, cammarota82, lehto84, lieb86,
bricmont96}. For instance, in his 1925 Ph.D. thesis, Ising showed that the
spin-spin correlation function\textbf{ }$\left\langle \sigma_{k}\sigma
_{l}\right\rangle $\textbf{ }in the one-dimensional Ising model with
Hamiltonian\textbf{ }$\mathcal{H}$\textbf{ }in the absence of an external
magnetic field decays exponentially with respect to the characteristic
length\textbf{ }$\xi$\textbf{ }\cite{ising}\textbf{,}%
\begin{equation}
\left\langle \sigma_{k}\sigma_{l}\right\rangle \overset{\text{def}}{=}\frac
{1}{\mathcal{Z}}\sum_{\left\{  \sigma_{i}\right\}  }\sigma_{k}\sigma
_{l}e^{-\beta\mathcal{H}}=e^{-\frac{\left\vert l-k\right\vert }{\xi}}\text{,}%
\end{equation}
where\textbf{ }$\left\vert l-k\right\vert $\textbf{ }denotes the distance
between sites\textbf{ }$k$\textbf{ }and\textbf{ }$l$\textbf{ }while\textbf{
}$\xi$\textbf{ }is defined as,%
\begin{equation}
\xi=\xi\left(  \beta\right)  \overset{\text{def}}{=}\frac{1}{\left\vert
\ln\left[  \tanh\left(  \beta J\right)  \right]  \right\vert }\text{,}%
\end{equation}
with\textbf{ }$J$\textbf{ }denoting the exchange coupling constant between
spins. Observe that $\xi$\textbf{ }diverges as\textbf{ }$\beta J$\textbf{
}approaches infinity. For\textbf{ }$T\neq0$\textbf{, }correlations decay
exponentially as a function of the correlation length. A short correlation
length means that distant spins are very weakly correlated. At high
temperatures,\textbf{ }$\beta J\ll1$\textbf{, }$\xi$\textbf{ }becomes
extremely short, and\textbf{ }$\left\langle \sigma_{k}\sigma_{l}\right\rangle
$\textbf{ }decays exponentially. To have a grasp of what high temperatures
means, let us assume that the exchange coupling constant between spins
is\textbf{ }$J\approx1$ eV and recalling that the Boltzmann constant
equals\textbf{ }$k_{B}\approx1.38\times10^{-23}$ J/K, an approximate estimate
of the temperature yields\textbf{ }$T\approx1\times10^{4}$ K $=10$ kK\textbf{.
}This order of temperature corresponds to the Fermi boiling point for a
valence electron, a temperature that for a metal is two orders of magnitude
above room temperature. Second, within statistical mechanics, the asymptotic
analysis of power-series expansions of expected values of observable
quantities is very important \cite{guttmann89}. In particular, it is the case
that series expansions agree well with high accuracy Monte Carlo simulations,
renormalization group results, and findings for exactly solvable models
\cite{wipf13}. In the high-temperature series expansion, the Boltzmann factor
is expanded in powers of the inverse temperature. For example, it can be shown
that the asymptotic high-temperature approximation of the nearest-neighbor
spin-spin correlation function\textbf{ }$\left\langle \sigma_{k}\sigma
_{l}\right\rangle $\textbf{ }in the two-dimensional Ising model in the absence
of an external magnetic field is given by \cite{guttmann89},%
\begin{equation}
\left\langle \sigma_{k}\sigma_{l}\right\rangle =\beta+\frac{5}{3}\beta
^{3}+\mathcal{O}\left(  \beta^{4}\right)  \text{.}%
\end{equation}
In what follows, we exploit the consequences of these two remarks in our own discussion.

\subsection{The formal comparison}

In the working hypothesis of extremely high temperatures, we observe that
Glauber's transition probability $\omega_{ij}$ is proportional to the square
root of the ratio between the stationary state probabilities of state $i$ and
state $j$. Specifically, the first-order expansion in the parameter $\beta$ of
Glauber's transition probability $\omega_{ij}$ in Eq. (\ref{omega}) leads to
the following approximate expression,%
\begin{equation}
\left[  \omega_{ij}\right]  _{\text{Glauber}}\overset{\beta\ll1}{\approx}%
\frac{1}{2}\Gamma e^{-\frac{\beta\Delta E}{2}}=\frac{1}{2}\Gamma\sqrt
{\frac{p_{j}}{p_{i}}}\text{,} \label{C1}%
\end{equation}
where, using Eq. (\ref{salva}), we find%
\begin{equation}
\frac{p_{j}}{p_{i}}=e^{-\beta\Delta E}\text{.} \label{ahah}%
\end{equation}
Furthermore, if we relax this working assumption and consider instead high
temperatures, we uncover that Glauber's transition probability $\omega_{ij}$
is no longer proportional to the square root of the ratio between the
stationary state probabilities of state $i$ and state $j$ and in this regime a
new approximate expression is obtained. Specifically, the second-order
expansion in the parameter $\beta$\ of Glauber's transition probability
$\omega_{ij}$\ in Eq. (\ref{hi}) yields,%
\begin{equation}
\left[  \omega_{ij}\right]  _{\text{Glauber}}=\frac{1}{2}\Gamma e^{-\frac
{\beta\Delta E}{2}}\left[  1-\frac{1}{2}\left(  \frac{\beta\Delta E}%
{2}\right)  ^{2}+O\left(  \beta^{4}\right)  \right]  \text{.} \label{glauby}%
\end{equation}
In what follows, in addition to considering very high temperature values $T$,
we also limit our analysis to energy difference values $\Delta E$\ in Eq
(\ref{glauby}) that belong to an interval of very small Lebesgue measure
proportional to $\delta E$\ that is centered at an energy difference value
$\Delta E_{\ast\text{ }}$that sets the energy scale of the stochastic Ising
model being considered. Specifically, our approximate analysis proceeds in the
following manner. We restrict our attention to energy difference values
$\Delta E$\ with $\Delta E_{\ast}-\delta E$\ $\leq\Delta E\leq\Delta E_{\ast
}+\delta E$\ and $0\leq\delta E\ll1$. In this case, the linear approximation
of $\Delta E^{2}$\ in the neighborhood of $\Delta E_{\ast}$\ is given by,%
\begin{equation}
\Delta E^{2}=\Delta E_{\ast}^{2}+2\Delta E_{\ast}\left(  \Delta E-\Delta
E_{\ast}\right)  +O\left(  \left\vert \Delta E-\Delta E_{\ast}\right\vert
^{2}\right)  \text{,} \label{linear}%
\end{equation}
that is,%
\begin{equation}
\Delta E^{2}\approx2\Delta E_{\ast}\Delta E-\Delta E_{\ast}^{2}\text{.}
\label{asymptotic}%
\end{equation}
Within this set of working hypotheses, combining Eqs. (\ref{glauby}) and
(\ref{asymptotic}) and recalling from Eqs. (\ref{hi}) and (\ref{g6}) that
$\Delta E\overset{\text{def}}{=}2\sigma_{i}J\left(  \sigma_{i-1}+\sigma
_{i+1}\right)  $, we obtain the following \emph{approximate} expression for
Glauber's transition probability%
\begin{equation}
\left[  \omega_{ij}\right]  _{\text{Glauber}}=\frac{1}{2}\Gamma e^{-\frac
{\beta\Delta E}{2}}\left[  1-\frac{1}{2}J\Delta E_{\ast}\beta^{2}\sigma
_{i}\left(  \sigma_{i-1}+\sigma_{i+1}\right)  +\frac{\Delta E_{\ast}^{2}}%
{8}\beta^{2}+O\left(  \beta^{4}\right)  \right]  \text{.}%
\end{equation}
Using Eq. (\ref{ahah}) and introducing the notations,%
\begin{equation}
\tilde{\Omega}\overset{\text{def}}{=}\frac{\Delta E_{\ast}^{2}}{8}\beta
^{2}\text{ and, }\tilde{\gamma}\overset{\text{def}}{=}J\Delta E_{\ast}%
\beta^{2}\text{,} \label{gt}%
\end{equation}
we finally determine that the new approximation for Glauber's $\omega_{ij}%
$\ becomes,%
\begin{equation}
\left[  \omega_{ij}\right]  _{\text{Glauber}}=\frac{1}{2}\Gamma\sqrt
{\frac{p_{j}}{p_{i}}}\left[  1-\frac{1}{2}\tilde{\gamma}\sigma_{i}\left(
\sigma_{i-1}+\sigma_{i+1}\right)  +\tilde{\Omega}+O\left(  \beta^{4}\right)
\right]  \text{,}%
\end{equation}
that is,%
\begin{equation}
\left[  \omega_{ij}\right]  _{\text{Glauber}}\overset{\beta^{2}\ll1}{\approx
}\frac{1}{2}\Gamma\sqrt{\frac{p_{j}}{p_{i}}}\left[  1-\frac{1}{2}\tilde
{\gamma}\sigma_{i}\left(  \sigma_{i-1}+\sigma_{i+1}\right)  +\tilde{\Omega
}\right]  \text{.} \label{C2}%
\end{equation}
We emphasize that the functional form obtained in Eq. (\ref{C2}) is not exact
and it is only \emph{approximately} valid in the limit of very high
temperatures $T$ and narrowly distributed energy changes $\Delta E$ in the
spin system being considered. Finally, we remark that the parameter
$\tilde{\gamma}$\ in Eq. (\ref{C2}) is a quadratic function of the parameter
$\beta$\ as evident from Eq. (\ref{gt}).\textbf{ }The approximate expressions
in Eqs. (\ref{C1}) and (\ref{C2}) will be compared with their analogs obtained
within the MaxCal platform. First, observe that in the absence of a
path-dependent dynamical information constraint, one sets the Lagrange
multiplier $\gamma$ in Eq. (\ref{fifi}) equal to zero. In this case, the
expression of the transition probability inferred by MaxCal reduces to,%
\begin{equation}
\left[  \omega_{ij}\right]  _{\text{Max-Cal}}=\mathcal{N}\sqrt{\frac{p_{j}%
}{p_{i}}}\text{.} \label{CC1}%
\end{equation}
Furthermore, if we assume as working hypothesis that the Lagrange multiplier
$\gamma$ in Eq. (\ref{fifi}) is nonvanishing but very small, that is,
$0\neq\gamma\ll1$, considering the first order series expansion in $\gamma$ of
the transition probability $\omega_{ij}$ in Eq. (\ref{fifi}), we obtain%
\begin{equation}
\left[  \omega_{ij}\right]  _{\text{Max-Cal}}=\mathcal{N}\sqrt{\frac{p_{j}%
}{p_{i}}}\left[  1-\frac{1}{2}\gamma\left(  \sigma_{ij}+\sigma_{ji}\right)
+O\left(  \gamma^{2}\right)  \right]  \text{,}%
\end{equation}
that is,%
\begin{equation}
\left[  \omega_{ij}\right]  _{\text{Max-Cal}}\overset{\gamma\ll1}{\approx
}\mathcal{N}\sqrt{\frac{p_{j}}{p_{i}}}\left[  1-\frac{1}{2}\gamma\left(
\sigma_{ij}+\sigma_{ji}\right)  \right]  \text{.} \label{CC2}%
\end{equation}
On the one hand, upon comparison of Eqs. (\ref{C1}) and (\ref{CC1}), we can
essentially identify the normalization factors $\mathcal{N}$ and $\Gamma$ and
conclude that the inferred transition probability inferred by the MaxCal
formalism in the absence of a path-dependent dynamical information constraint
exhibits the same square-root dependence that appears in Glauber's approximate
expression (first-order expansion in $\beta$, extremely high temperatures) of
the transition probability in the limiting case of extremely high temperature
values. On the other hand, comparison of Eqs. (\ref{C2}) and (\ref{CC2}), in
addition to identification of the normalization factors $\mathcal{N}$ and
$\Gamma$ (more specifically\textbf{, }$\mathcal{N}\leftrightarrow
\frac{1+\tilde{\Omega}}{2}\Gamma$), allows to link the quantity $\gamma$ in
Eq. (\ref{fifi}) to the quantity $\tilde{\gamma}$ in\ Eq. (\ref{gt}) (more
specifically\textbf{, }$\gamma\leftrightarrow\frac{1}{1+\tilde{\Omega}}%
\tilde{\gamma}$) and exploit the relation $\sigma_{ij}+\sigma_{ji}=\sigma
_{i}\left(  \sigma_{i-1}+\sigma_{i+1}\right)  $ in the working hypothesis of
nearest-neighbor interactions of each spin\textbf{ }$\sigma_{i}$ with pairs of
spins\textbf{ }$\sigma_{j}$\textbf{ }with\textbf{ }$j\in\left\{
i\pm1\right\}  $\textbf{. }We then uncover that the MaxCal formalism infers an
approximate expression (first-order expansion in $\gamma$) of the transition
probability that is functionally identical to the approximate expression
(second-order expansion in $\beta$, high temperatures) obtained from Glauber's
analysis. The fact that we have equated a relation obtained from a first-order
expansion in $\gamma$ (MaxCal, Eq. (\ref{CC2})) with a relation derived from a
second-order expansion in $\beta$ (Glauber, Eq. (\ref{C2}))\textbf{
}is\textbf{ }remarkably consistent since we have identified the MaxCal
Lagrange multiplier $\gamma$ with $\tilde{\gamma}=\tilde{\gamma}\left(
\beta\right)  \propto\beta^{2}$ in Eq. (\ref{gt}).

We remark that the MaxCal algorithm allows us to make plausible inferences but
not logical deductions. Such inferences rely on the nature of the chosen
information constraints used in the algorithm. The validation of this type of
modeling scheme can be checked only \emph{a posteriori}. If discrepancies
between the inferred predictions and experimental observations are recorded, a
different set of information constraints has to be chosen. In our analysis, we
do not recover the exact functional form of Glauber's transition probability.
Our correspondence between the MaxCal and Glauber's solutions is only
\emph{approximately} valid in the limit of very high temperatures and narrowly
distributed energy changes in the spin system being investigated. A more
refined attempt to recover the exact expression would also require the clever
introduction of some sort of information constraint that captures the \emph{ad
hoc} assumption employed by Glauber, that is, the symmetric dependence of the
transition probability on the two neighboring spins $\sigma_{i-1}$ and
$\sigma_{i+1}$ as well as $\sigma_{i}$.

\section{Conclusive Remarks}

In this article, we employed the MaxCal variational principle as an inference
algorithm used to predict dynamical properties of complex nonequilibrium
stationary statistical systems in the presence of incomplete information.
Specifically, we maximized the path entropy over discrete time step
trajectories subject to normalization, stationarity, and detailed balance
constraints together with a path-dependent dynamical information constraint
reflecting a suitably chosen average global behavior of the complex system.
Furthermore, we considered a path-dependent information constraint defined in
terms of the average of the product of two neighboring spin values $\sigma
_{i}\left(  t\right)  $ and $\sigma_{j}\left(  t\right)  $ as specified in Eq.
(\ref{c4}). \ A general expression for the transition probability associated
to the stationary random Markov processes describing the nonequilibrium
stationary system was computed and reported in Eq. (\ref{fifi}),%
\begin{equation}
\left[  \omega_{ij}\right]  _{\text{Max-Cal}}=\mathcal{N}\sqrt{\frac{p_{j}%
}{p_{i}}}\cosh\left[  \frac{\gamma}{2}\left(  \sigma_{ij}+\sigma_{ji}\right)
\right]  \left\{  1-\tanh\left[  \frac{\gamma}{2}\left(  \sigma_{ij}%
+\sigma_{ji}\right)  \right]  \right\}  \text{.} \label{xxx}%
\end{equation}
The expression in Eq. (\ref{xxx}) was then compared to the well-known Glauber
hyperbolic tangent rule for the transition probability that characterizes the
stochastic Ising model as recast by us in Eq. (\ref{omega}),%
\begin{equation}
\left[  \omega_{ij}\right]  _{\text{Glauber}}=\Gamma\frac{e^{-\frac
{\beta\Delta E}{2}}}{e^{\frac{\beta\Delta E}{2}}+e^{-\frac{\beta\Delta E}{2}}%
}\text{.} \label{xxxx}%
\end{equation}
The comparison was presented in two limiting cases. In the first case, we
uncovered that in the absence of a path-dependent dynamical information
constraint, $\omega_{ij}$ in Eq. (\ref{xxx}) exhibits the same square-root
dependence that appears in Glauber's approximate expression (first-order
expansion in $\beta$, extremely high temperatures) of the transition
probability $\omega_{ij}$ in Eq. (\ref{xxxx}) in the limiting case of
extremely high temperature values. This first comparison was performed by
considering Eqs. (\ref{C1}) and (\ref{CC1}). In the second case, we uncovered
that in the presence of a path-dependent dynamical information constraint the
MaxCal formalism infers an approximate expression (first-order expansion in
$\gamma$) of the transition probability $\omega_{ij}$ in Eq. (\ref{xxx}) whose
functional structure is identical to that of the approximate expression
(second-order expansion in $\beta$, high temperatures) obtained from Glauber's
$\omega_{ij}$ in Eq. (\ref{xxxx}). This second comparison was performed by
considering Eqs. (\ref{C2}) and (\ref{CC2}).

In summary, the main findings of our scientific activity reported in this
manuscript can be outlined as follows:

\begin{itemize}
\item We established a quantitative link between the MaxCal formalism
\cite{jaynes80, jaynes85, karpov1, karpov2} and the stochastic Ising model as
originally presented by Glauber \cite{glauber63}. Specifically, a connection
between the transition probability inferred by MaxCal and Glauber's hyperbolic
tangent rule is proposed in two limiting scenarios.

\item We advanced the line of research based on the MaxCal inference
algorithm, especially the one advocated by Dixit \cite{dixit15A} by extending
the applicability of his use of the MaxCal formalism to a dynamical constraint
in the form of a path ensemble average of the product of two neighboring spin values.

\item We significantly elaborated on the very intriguing preliminary remark
concerning the relation between MaxCal and Glauber's dynamics as recently
reported in Ref. \cite{dixit15B}. Our elaboration led to an important
advancement in the conceptual understanding of the above mentioned preliminary consideration.
\end{itemize}

We think our work presented herein is a valid addition to our continuing
effort of providing a unifying theoretical framework of statistical mechanical
nature for describing and understanding complex systems of arbitrary nature in
the presence of incomplete information \cite{cafaro16}. As pointed out by
Feynman in his Nobel lecture \cite{feynman72} , using different mathematical
approaches for describing the same physical result can provide a better
starting point for subsequent reasoning. This statement seems to be especially
well-suited for the connection between the MaxCal formalism and the
conventional nonequilibrium statistical mechanical thinking. MaxCal could
potentially be used to understand what is not yet understood. In light of
these considerations, it is our sincere hope that other scientists will find
our work presented here relevant and worthy of further refinement.

\begin{acknowledgments}
C. C. is grateful to Adom Giffin and Domenico Felice for discussions on the
MaxCal formalism and the Glauber dynamics, respectively. Constructive
criticism from two anonymous referees leading to an improved version of this
manuscript are sincerely acknowledged by the authors. Finally, C. C.
acknowledges the hospitality of the Air Force Research Laboratory (AFRL) where
part of his contribution to this work was completed.
\end{acknowledgments}

\bigskip

\bigskip


\begin{thebibliography}{99}                                                                                               %


\bibitem {gibbs74}J. W. Gibbs, \emph{On the equilibrium of heterogeneous
substances}, Transactions of the Connecticut Academy of Arts and Sciences,
Vol. 3 (1874-78), pp. 108-248 and 343-524.

\bibitem {shannon48}C. E.\ Shannon, \emph{A mathematical theory of
communication}, The Bell System Technical Journal\textbf{\ 27}, 379 (1948).

\bibitem {jaynes57A}E. T. Jaynes,\emph{\ Information theory and statistical
mechanics}, Phys. Rev. \textbf{106}, 620 (1957).

\bibitem {jaynes57B}E. T. Jaynes,\emph{\ Information theory and statistical
mechanics. II}, Phys. Rev. \textbf{108}, 171 (1957).

\bibitem {jaynes80}E. T. Jaynes, \emph{The minimum entropy production
principle}, Ann. Rev. Phys. Chem. \textbf{31}, 579 (1980).

\bibitem {jaynes85}E. T. Jaynes, \emph{Macroscopic prediction}, in Complex
Systems- Operational Approaches in Neurobiology, Physics, and Computers, H.
Haken, Ed.; Springer-Verlag, Berlin, pp. 254 (1985).

\bibitem {jaynes86}E. T. Jaynes, \emph{Predicitive statistical mechanics}, in
Frontiers of Nonequilibrium Statistical Physics, G. T. Moore and M.\emph{ }O.
Scully, Eds.; Plenum Press, New York, p. 33 (1986).

\bibitem {rondoni14}S. Chibbaro, L. Rondoni, and A. Vulpiani, \emph{On the
foundations of statistical mechanics: ergodicity, many degrees of freedom and
inference}, Comm. Theor. Phys. \textbf{62}, 469 (2014).

\bibitem {dewar09}R. C. Dewar, \emph{Maximum entropy production as an
inference algorithm that translates physical assumptions into macroscopic
predictions: don't shoot the messenger}, Entropy \textbf{11}, 931 (2009).

\bibitem {gull78}S. F. Gull and G. J. Daniell, \emph{Image reconstruction from
incomplete and noisy data}, Nature \textbf{272}, 686 (1978).

\bibitem {skilling84}J. Skilling and R. K. Bryan, \emph{Maximum entropy image
reconstruction: general algorithm}, Mon. Not. R. Astr. Soc. \textbf{211}, 111 (1984).

\bibitem {minerbo79}G. Minerbo, \emph{A maximum entropy algorithm for
reconstructing a source from projected data}, Comp. Gaph. Im. Proc.
\textbf{10}, 48 (1979).

\bibitem {collins82}D. M. Collins, \emph{Electron density images from
imperfect data by iterative entropy maximization}, Nature \textbf{298}, 49 (1982).

\bibitem {bryan83}R. K. Bryan, M. Bansal, W. Folkhard, C. Nave, and D. A.
Marvin, \emph{Maximum-entropy calculation of the electron density at }%
$4A$\emph{\ resolution of }$Pfl$\emph{\ filamentous bacteriophage}, Proc. Nat.
Acad. Sci. \textbf{80}, 4728 (1983).

\bibitem {sibisi84}S. Sibisi, J.\ Skilling, R. G. Brereton,\ E.\ D. Lane, and
J. Staunton, \emph{Maximum entropy signal processing in practical NMR
spectroscopy}, Nature \textbf{311}, 446 (1984).

\bibitem {mora15}T. Mora, S. Deny, and O. Marre, \emph{Dynamical criticality
in the collective activity of a population of retinal neurons}, Phys. Rev.
Lett. \textbf{114}, 078105 (2015).

\bibitem {carlopd07}C. Cafaro and S. A. Ali, \emph{Jacobi fields on
statistical manifolds of negative curvature}, Physica \textbf{D234}, 70 (2007).

\bibitem {carlochaos}C. Cafaro, \emph{Works on an information
geometrodynamical approach to chaos}, Chaos, Solitons \& Fractals \textbf{41},
886 (2009).

\bibitem {carlopd11}C. Cafaro and S. Mancini, \emph{Quantifying the complexity
of geodesic paths on curved statistical manifolds through information
geometric entropies and Jacobi fields}, Physica \textbf{D240}, 607 (2011).

\bibitem {henry16}G. Henry and D. Rodriguez, \emph{On the instability of two
entropic dynamical models}, Chaos,\ Solitons \& Fractals \textbf{91}, 604 (2016).

\bibitem {carlomplb08}C. Cafaro, \emph{Information geometry, inference methods
and chaotic energy levels statistics}, Mod. Phys. Lett. \textbf{B22}, 1879 (2008).

\bibitem {carlopa08}C. Cafaro and S. A. Ali, \emph{Can chaotic quantum energy
levels statistics be characterized using information geometry and inference
methods?}, Physica \textbf{A387}, 6876 (2008).

\bibitem {carlopla}D.-H. Kim, S. A. Ali, C. Cafaro\textbf{,} and S. Mancini,
\emph{Information geometric modeling of scattering induced quantum
entanglement}, Phys. Lett. \textbf{A375}, 2868 (2011).

\bibitem {carlopa}D.-H. Kim, S. A. Ali, C. Cafaro\textbf{,} and S. Mancini,
\emph{Information geometry of quantum entangled wave-packets}, Physica
\textbf{A391}, 4517 (2012).

\bibitem {carlo16}A. Giffin, C. Cafaro, and S. A. Ali, \emph{Application of
the maximum relative entropy method to the physics of ferromagnetic
materials}, Physica \textbf{A455}, 11 (2016).

\bibitem {ghosh06}K. Ghosh, K. A.\ Dill, M. M. Inamdar, E. Seitaridou, and R.
Phillips, \emph{Teaching the principles of statistical dynamics}, Am. J. Phys.
\textbf{74}, 123 (2006).

\bibitem {seita07}E. Seitaridou, M. M. Inamdar, R. Phillips, K. Ghosh, and K.
Dill, \emph{Measuring flux distributions for diffusion in the small-numbers
limit}, J. Phys. Chem. \textbf{B111}, 2288 (2007).

\bibitem {wu09}D.\ Wu, K. Ghosh, M. Inamdar, H. L. Lee, S. Fraser, K. Dill,
and R. Phillips, \emph{Trajectory approach to two-state kinetics of single
particles on sculpted energy landscapes}, Phys. Rev. Lett. \textbf{103},
050603 (2009).

\bibitem {cavagna14}A. Cavagna, I. Giardina, F. Ginelli, T. Mora, S. Piovani,
R. Tavarone, and A. M. Walczak, \emph{Dynamical maximum entropy approach to
flocking}, Phys. Rev. \textbf{E89}, 042707 (2014).

\bibitem {jtd94}J. T. Dougherty, \emph{Foundations of non-equilibrium
statistical mechanics}, Phil. Trans. R.\ Soc. Lond. \textbf{A346}, 259 (1994).

\bibitem {dewar03}R. Dewar, \emph{Information theory explanation of the
fluctuation theorem, maximum entropy production and self-organized criticality
in non-equilibrium stationary states}, J. Phys. \textbf{A}: Math. Gen.
\textbf{36}, 631 (2003).

\bibitem {gonzalez13}D. Gonzalez, S. Davis, and G. Gutierrez, \emph{Newtonian
dynamics from the principle of maximum caliber}, Found. Phys. \textbf{44}, 923 (2014).

\bibitem {davis15}S. Davis and D. Gonzalez, \emph{Hamiltonian formalism and
path entropy maximization}, J. Phys. \textbf{A}: Math. Theor. \textbf{48},
425003 (2015).

\bibitem {stock08}G. Stock, K. Ghosh, and K. A. Dill, \emph{Maximum caliber: A
variational approach applied to two-state dynamics}, J. Chem. Phys.
\textbf{128}, 194102 (2008).

\bibitem {lee12}J. Lee and S. Presse, \emph{A derivation of the master
equation from path entropy maximization}, J. Chem. Phys. \textbf{137}, 074103 (2012).

\bibitem {ge12}H. Ge, S. Presse, K. Ghosh, and K. A. Dill, \emph{Markov
processes follow the principle of maximum caliber}, J. Chem. Phys.
\textbf{136}, 064108 (2012).

\bibitem {otten10}M. Otten and G. Stock, \emph{Maximum caliber inference of
nonequilibrium processes}, J. Chem. Phys. \textbf{133}, 034119 (2010).

\bibitem {hazoglou15}M. J. Hazoglou, V. Walther, P. D. Dixit, and K. A. Dill,
\emph{Maximum caliber is a general variational principle for nonequilibrium
statistical mechanics}, J. Chem. Phys. \textbf{143}, 051104 (2015).

\bibitem {presse13}S. Presse, K. Ghosh, J. Lee, and K. A. Dill,
\emph{Principles of maximum entropy and maximum caliber in statistical
physics}, Rev. Mod. Phys. \textbf{85}, 1115 (2013).

\bibitem {karpov1}A. A. Filyukov and V. Y. Karpov, \emph{Method of the most
probable path of evolution in the theory of stationary irreversible
processes}, J.\ Engineering Physics \textbf{13}, 416 (1967).

\bibitem {karpov2}A. A. Filyukov and V. Y. Karpov, \emph{Description of steady
transport processes by the method of the most probable path of evolution},
J.\ Engineering Physics \textbf{13}, 326 (1967).

\bibitem {karpov3}A. A. Filyukov, \emph{Compatibility property of steady
systems}, J.\ Engineering Physics \textbf{14}, 429 (1968).

\bibitem {dixit14}P. D. Dixit and K. A. Dill, \emph{Inferring microscopic
kinetic rates from stationary state distributions}, Journal of Chemical Theory
and Computation \textbf{10}, 3002 (2014).

\bibitem {dixit15A}P. D. Dixit, \emph{Stationary properties of maximum entropy
random walks}, Phys. Rev. \textbf{E92}, 042149 (2015).

\bibitem {dixit15B}P. D. Dixit, A. Jain, G. Stock, and K. A. Dill,
\emph{Inferring transition rates on networks with incomplete knowledge},
arXiv:q-bio.MN/1504.01277 (2015).

\bibitem {huangbook}K. Huang, \emph{Statistical Mechanics}, John Wiley \&
Sons, Inc. (1987).

\bibitem {huang73}H. W. Huang, \emph{Time-dependent statistics of the Ising
model in a magnetic field}, Phys. Rev. \textbf{A8}, 2553 (1973).

\bibitem {ising}E. Ising, Beitrag zur theorie des ferromagnetismus,
Zeitschrift fur Physik \textbf{31}, 253 (1925).

\bibitem {glauber63}R. J. Glauber, \emph{Time-dependent statistics of the
Ising model}, J. Math. Phys. \textbf{4}, 294 (1963).

\bibitem {penrose91}O. Penrose, \emph{A mean-field equation of motion for the
dynamic Ising model}, J. Stat. Phys. \textbf{63}, 975 (1991).

\bibitem {binder05}D. P. Landau and K. Binder, \emph{A Guide to Monte Carlo
Simulations in Statistical Physics}, Cambridge University Press (2005).

\bibitem {janke12}W. Janke, \emph{Monte Carlo Simulations in Statistical
Physics- From Basic Principles to Advanced Applications}. In: Order, Disorder
and Criticality: Advanced Problems of Phase Transition Theory, Vol. 3, pp.
93-166, edited by Yu. Holovatch, World Scientific Publishing, Singapore (2012).

\bibitem {kampen81}N. G. van Kampen, \emph{Stochastic Processes in Physics and
Chemistry}, North Holland, Amsterdam (1981).

\bibitem {cover}T. M. Cover and J. A. Thomas, \emph{Elements of Information
Theory}, John Wiley \& Sons, Inc., New Jersey (2006).

\bibitem {gross79}L. Gross, \emph{Decay of correlations in classical lattice
models at high temperature}, Comm. Math. Phys. \textbf{68}, 9 (1979).

\bibitem {cammarota82}C. Cammarota, \emph{Decay of correlations for infinite
range interactions in unbounded spin systems}, Comm. Math. Phys. \textbf{85},
517 (1982).

\bibitem {lehto84}M. Lehto, H. B. Nielsen, and M. Ninomiya, \emph{A
correlation decay theorem at high temperature}, Comm. Math. Phys. \textbf{93},
483 (1984).

\bibitem {lieb86}T. Kennedy and E. H. Lieb, \emph{An itinerant electron model
with crystalline or magnetic long range order}, Physica \textbf{A138}, 320 (1986).

\bibitem {bricmont96}J. Bricmont and A. Kupiainen, \emph{High temperature
expansions and dynamical systems}, Comm. Math. Phys. \textbf{178}, 703 (1996).

\bibitem {guttmann89}A. J. Guttmann, \emph{Asymptotic analysis of power-series
expansions}, 1-234. In Phase Transitions and Critical Phenomena, Vol. 3,
edited by C. Domb and J. L. Lebowitz, Academic Press, London (1989).

\bibitem {wipf13}A. Wipf, \emph{High-temperature and low-temperature
expansions}, Lecture Notes in Physics \textbf{864}, 173 (2013).

\bibitem {cafaro16}C. Cafaro, S. A. Ali, and A. Giffin, \emph{Thermodynamic
aspects of information transfer in complex dynamical systems}, Phys.
Rev.\ \textbf{E93}, 022114 (2016).

\bibitem {feynman72}R. P. Feynman, \emph{The development of the space-time
view of quantum electrodynamics}, Nobel Lectures, Physics 1963-1970, Elsevier
Publishing Company, Amsterdam (1972).
\end{thebibliography}
\end{document}